# Towards twin-free molecular beam epitaxy of 2D chalcogenides explained by stronger interlayer van der Waals coupling

Wouter Mortelmans[1,2,*], Karel De Smet[2,3], Ruishen Meng[4], Michel Houssa[4], Stefan De Gendt[2,3], Marc Heyns[1,2] and Clement Merckling[1,2,*]

[1]KU Leuven, Department of Materials Engineering, Kasteelpark Arenberg 44, 3001 Leuven, Belgium
[2]Imec, Kapeldreef 75, 3001 Leuven, Belgium
[3]KU Leuven, Department of Chemistry, Celestijnenlaan 200f, 3001 Leuven, Belgium
[4]KU Leuven, Department of Physics and Astronomy, Celestijnenlaan 200d, 3001 Leuven, Belgium
[*]Corresponding authors: wouter.mortelmans@imec.be; clement.merckling@imec.be

**ABSTRACT:**

Defect-free epitaxial growth of 2D materials is one of the holy grails for a successful integration of van der Waals (vdW) materials in the semiconductor industry. The large-area (quasi-)vdW epitaxy of layered 2D chalcogenides is consequently carefully being researched since these materials hold very promising properties for future nanoelectronic applications. The formation of defects such as stacking faults like 60º twins and consequently 60º grain boundaries is still of major concern for the defect-free epitaxial growth of 2D chalcogenides. Although growth strategies to overcome the occurrence of these defects are currently being considered, more fundamental understanding on the origin of these defects at the initial stages of the growth is highly essential. Therefore this work focuses on the understanding of 60º twin formation in (quasi-)vdW epitaxy of 2D chalcogenides relying on systematic molecular beam epitaxy (MBE) experiments supported by density functional theory (DFT) calculations. The MBE experiments reveal the striking difference in 60º twin formation between $WSe_2$ and $Bi_2Se_3$ in both quasi-vdW heteroepitaxy and vdW homoepitaxy, which from our DFT calculations links to the difference in interlayer vdW coupling strength. The stronger interlayer vdW coupling in $Bi_2Se_3$ compared to $WSe_2$ results in a striking enhanced control on twin formation and hence shows significantly more promise for defect-free epitaxial integration. This interesting aspect of (quasi-)vdW epitaxy reveals that the strength of interlayer vdW coupling is key for functional 2D materials and opens perspectives for other vdW materials sharing strong interlayer interactions.

**KEYWORDS:**

van der Waals epitaxy, molecular beam epitaxy, density functional theory, 2D chalcogenides, stacking faults.

The discovery of graphene and its unique transport properties in 2004 by K. Novoselov and A. Geim has boosted recent interests in a broad variety of 2D materials[1,2]. Layered chalcogenides are, in this framework, a promising family of van der Waals (vdW) materials[3–5]. Important candidates are the transition metal dichalcogenides (TMDs) having the chemical form of $MX_2$ and the topological insulators (TIs) having the chemical form of group-$V_2VI_3$. TMD materials such as $WSe_2$, $MoS_2$, etc. are highly interesting for opto- and nanoelectronic applications thanks to their semiconducting properties and direct bandgap at monolayer (ML) thickness[6–11]. TI materials such as $Bi_2Se_3$, $Sb_2Te_3$, ... and alloys are most promising for their topology owing to the strong spin-orbit coupling and band inversion enabling new states of quantum matter in which surface states from the bulk insulating gap are spin-polarized and protected by time-reversal symmetry[12,13].

The large-area integration of 2D chalcogenides is of crucial importance for these materials to become a mature option and enable industry-compatible devices[14]. The growth of vdW materials through the process of epitaxy is one of the most promising approaches to meet the demanding requirements of single-crystalline quality, large-area uniformity, and large-scale throughput[15–17]. Therefore, both quasi-vdW (2D on 3D) and vdW (2D on 2D) epitaxy of layered chalcogenides are extensively being research in the literature[18–20]. One of the major concerns in (quasi-)vdW epitaxy of these materials is the systematic formation of stacking faults like 60° twins, as observed in either molecular beam epitaxy (MBE) (quasi-vdW[21–29] and vdW[21,30–36]), metalorganic vapor phase epitaxy (MOVPE) (quasi-vdW[21,37–42] and vdW[21,43,44]) and chemical vapor epitaxy (CVE) (quasi-vdW[45–47] and vdW[48–53]). To mitigate the formation of 60° twin defects, several approaches are being reported that rely on optimized growth conditions[54–56], buffer layer growth[57], or the introduction of a 3D aspect in the growth surface like surface roughness[58] or surface step edges[59].

However, to date, a more fundamental understanding on the formation of 60° twin defects in (quasi-)vdW epitaxy of layered chalcogenides is highly required. Therefore, a systematic comparative study focusing on the formation of twin defects is performed in this work for the epitaxy of various 2D chalcogenides using the MBE growth technique. The 2D chalcogenides that are studied are $WSe_2$ from the $MX_2$ family and $Bi_2Se_3$ from the group-$V_2VI_3$ family. The epitaxial processes include $WSe_2$ and $Bi_2Se_3$ quasi-vdW heteroepitaxy on on-axis c-plane sapphire substrates and $WSe_2$ and $Bi_2Se_3$ vdW homoepitaxy on respectively $WSe_2(0001)$ and $Bi_2Se_3(0001)$ exfoliated flakes. The similarities and differences of the various epitaxy processes are presented and discussed in the frame to shed more light on the fundamental aspect of 60° twin formation.

The experimental methodology that is applied for the 2D chalcogenide MBE processes is illustrated using simplified schematics in Figure 1. The (1x1) reconstructed sapphire surfaces are obtained by thermal annealing as reported and characterized previously[23]. The virtual $WSe_2(0001)$ and $Bi_2Se_3(0001)$ substrates are fabricated relying on mechanical exfoliation on silicon substrates as reported before[33,34]. The epitaxies are performed using plasma-assisted (PA-)MBE with $H_2X$ radio frequency (RF) plasma sources[60] and electron-beam evaporation of elemental W transition metals (Figure 1a) and thermal evaporation of elemental Bi metals (Figure 1b). For the $WSe_2$ compound, the growths occur at a temperature of 450 °C with low growth rates of ~ 0.1-1.3 ML.h$^{-1}$, completely driven by the W evaporation flux. The growths of the $Bi_2Se_3$ compound occur at a lower temperature of 160 °C with higher growth rates of ~ 5-12 ML.h$^{-1}$, and similarly, completely driven by the Bi evaporation flux. The $H_2Se$ flux is set to a total and maximal pressure of ~ 2.0 x 10$^{-5}$ Torr in the RF plasma source.

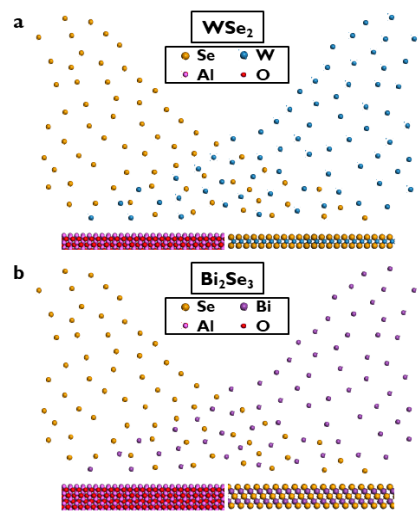

**Figure 1:** Simplified schematic illustration of the experimental setup used to study the MBE (quasi-)vdW epitaxy of $WSe_2$ and $Bi_2Se_3$ on sapphire and on respectively $WSe_2$ and $Bi_2Se_3$ surfaces. a) $WSe_2$ PA-MBE method relying on the electron-beam evaporation of elemental W and $H_2Se$ plasma. b) $Bi_2Se_3$ PA-MBE method relying on the thermal evaporation of elemental Bi and $H_2Se$ plasma.

The quasi-vdW heteroepitaxies of $WSe_2$ and $Bi_2Se_3$ on the sapphire surfaces are presented in Figure 2. They represent 1 ML of $WSe_2$ (Figures 2a-b) and 1 ML of $Bi_2Se_3$ (Figures 2c-d) on (1x1) c-plane sapphire substrates. This single-layer thickness is chosen to maximize the amount of quasi-vdW heteroepitaxy while avoiding the vdW homoepitaxy of the 2$^{nd}$ ML on the 1$^{st}$ ML.

The polar RHEED characterization of the $WSe_2$ quasi-vdW heteroepitaxy clearly reveals the in-plane epitaxial registry of the $WSe_2$ crystals with the underlying $Al_2O_3$ surface (Figure 2a). The epitaxial relation is highlighted using yellow and blue diamonds for respectively WSe$_2$ and Al$_2$O$_3$ and is characterized as $[11\bar{2}0]$WSe$_2$(0001)// $[11\bar{2}0]$Al$_2$O$_3$(0001). This is a similar epitaxial relationship as previously reported for the growths of WSe$_2$/MoS$_2$ on various reconstructed sapphire surfaces[21,23,37]. However, the identical $(0\bar{1})$ and $(01)$ diffraction streaks observed from the diffraction patterns uncovers an important limitation of the quasi-vdW epitaxy experiment. In Figure 2b, the RHEED pattern in the WSe$_2\langle 11\bar{2}0\rangle$ direction is presented where several intensity line profiles are extracted from various '$k_z$' positions that give information about the out-of-plane ordering of the grown 2D crystal planes[61]. The equivalent intensities of the $(0\bar{1})$ and $(01)$ diffraction streaks confirm the absence of the expected 3-fold periodic stacking hence the abundant presence of 60° twins[26]. This results from the lack of a preferred stacking in the $[11\bar{2}0]$WSe$_2$(0001)// $[11\bar{2}0]$Al$_2$O$_3$(0001) registry and consequently results in a high defect density of 60° grain boundaries impacting device performances[62,63].

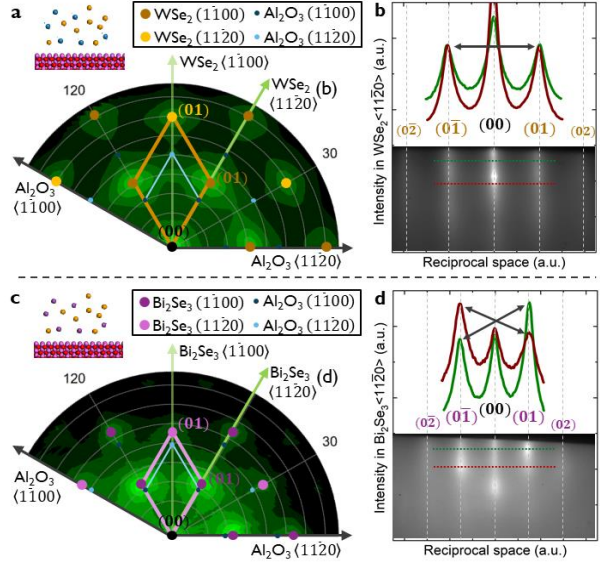

**Figure 2:** Epitaxial registry and preferred stacking in MBE quasi-vdW heteroepitaxy of 2D chalcogenides. The top panel (a-b) corresponds to the WSe$_2$ quasi-vdW heteroepitaxy on (1×1) sapphire. The bottom panel (c-d) corresponds to the Bi$_2$Se$_3$ quasi-vdW heteroepitaxy on (1×1) sapphire. a,c) Azimuthal RHEED scans overlaid with yellow (WSe$_2$) and purple (Bi$_2$Se$_3$) dots representing the diffractions of respectively WSe$_2$ and Bi$_2$Se$_3$ and with blue dots representing the ones of α-Al$_2$O$_3$. The diamond reveals the in-plane alignment of the grown vdW crystals with respect to the Al$_2$O$_3$ surface underneath. b,d) RHEED intensity line profiles and pattern in respectively the WSe$_2\langle 11\bar{2}0\rangle$ and the Bi$_2$Se$_3\langle 11\bar{2}0\rangle$ direction. The equivalent $(0\bar{1})$ and $(01)$ diffraction streaks of the WSe$_2$ and the inequivalent $(0\bar{1})$ and $(01)$ diffraction streaks of the Bi$_2$Se$_3$ demonstrate respectively the absence and presence of the 3-fold in-plane characteristic hence the preferred 3-fold symmetric stacking in the quasi-vdW heteroepitaxy on sapphire.

Remarkably, in $Bi_2Se_3$ quasi-vdW heteroepitaxy, a striking difference with respect to $WSe_2$ quasi-vdW heteroepitaxy is observed. This is obtained from the polar RHEED characterization presented in Figures 2c-d. The epitaxial relation of the $Bi_2Se_3$ with the (1x1) sapphire surface is similar as for the case of $WSe_2$: $[11\bar{2}0]Bi_2Se_3(0001)// [11\bar{2}0]Al_2O_3(0001)$ (Figure 2c, purple and blue diamonds for respectively $Bi_2Se_3$ and $Al_2O_3$). However, the stacking preference and hence the occurrence of 60° twins (and 60° grain boundaries) is strikingly dissimilar. This is corroborated from the inequivalent $(0\bar{1})$ and $(01)$ diffraction streaks of the $Bi_2Se_3$ quasi-vdW heteroepitaxy as observed from the $Bi_2Se_3\langle11\bar{2}0\rangle$ RHEED pattern (Figure 2d). The observation of these inequivalent streaks is in agreement with the 3-fold in-plane rotational symmetry of the $Bi_2Se_3$ crystal structure, and confirms the preferred and unique stacking of $Bi_2Se_3$ on sapphire and hence the reduced formation of 60° twins[64]. Surprisingly, $Bi_2Se_3$ quasi-vdW heteroepitaxy is less prone to stacking fault formation compared to $WSe_2$ (and in general TMDs), despite the equivalent in-plane crystal structure symmetry and presence of vdW gap in both compounds. Consequently, $Bi_2Se_3$ shows significantly more promise for defect-free epitaxial integration[65].

To further explore these interesting aspects, a study is presented based on the vdW homoepitaxy of the highlighted 2D chalcogenides. The vdW homoepitaxy experiments represent ~1/3 ML of $WSe_2$ and $Bi_2Se_3$ grown on respectively exfoliated $WSe_2(0001)$ and $Bi_2Se_3(0001)$ flakes, obtained by a reduction of the growth rate compared to the experiments performed on sapphire. Such limited thickness is preferred here, to enable the identification of the individual grown 2D chalcogenide crystals before coalescence and to avoid the onset of multilayer growth.

The $WSe_2$ vdW homoepitaxy is identified by a high density of characteristic triangular grains with crystal sizes up to ~50 nm. This is observed from the AFM image presented in Figure 3a. The algorithmic analysis of the $WSe_2$ crystals nucleated and grown on the $WSe_2(0001)$ surface enables to qualify the epitaxial relation and to reveal the presence of a preferred stacking[33]. The distribution of the relative azimuthal in-plane orientation of the analyzed $WSe_2$ triangular grains is presented in Figure 3b. This distribution with a 60° difference between both consecutive peaks clearly reveals the 6-fold in-plane periodicity which is reported to result from the inability to control the bilayer stacking phase[33]. Consequently, both the 2H and 3R stacking phases are simultaneously present in the vdW homoepitaxy confirming the presence of a high density of 60° twins and 60° grain boundaries upon closure of the ML. Hence, in both vdW homoepitaxy and quasi-vdW heteroepitaxy, the WSe$_2$ compound systematically yields severe formation of stacking faults at the early stage of the growth which is emphasized to be very challenging to control.

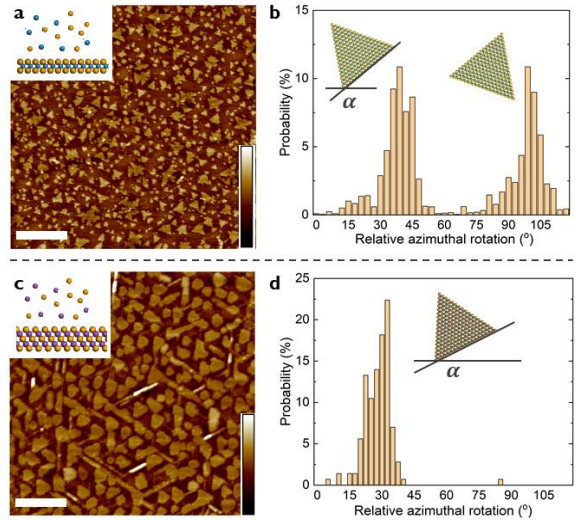

**Figure 3:** Epitaxial registry and preferred stacking in MBE vdW homoepitaxy of 2D chalcogenides. The top panel (a-b) corresponds to the WSe$_2$ vdW homoepitaxy on WSe$_2$(0001) surfaces. The bottom panel (c-d) corresponds to the Bi$_2$Se$_3$ vdW homoepitaxy on Bi$_2$Se$_3$(0001) surfaces. a,c) AFM images of respectively the WSe$_2$ and Bi$_2$Se$_3$ vdW homoepitaxy. b,d) Analyses of the AFM images in (a,c) highlighting the relative azimuthal in-plane distribution of the nucleation and grown crystals in respectively the WSe$_2$ and Bi$_2$Se$_3$ vdW homoepitaxy. The 6-fold periodicity of the WSe$_2$ crystals and the 3-fold periodicity of the Bi$_2$Se$_3$ crystals demonstrate respectively the absence and presence of a preferred stacking in the vdW homoepitaxial registry.

This observation is - once more - in striking difference with the Bi$_2$Se$_3$ compound. The AFM characterization and crystal analysis of the Bi$_2$Se$_3$ vdW homoepitaxy is presented in respectively Figures 3c and 3d. The Bi$_2$Se$_3$ vdW homoepitaxy yields characteristic equilateral triangles having a crystal grain size up to ~100 nm, and a 3-fold periodic in-plane alignment of the nucleated and grown crystals as in agreement with the symmetry of the Bi$_2$Se$_3$ crystal structure. The larger grain size is linked with the larger vapor pressure of the elemental bismuth, since adatom diffusion is previously reported to correlate with vapor pressure in vdW epitaxy of TMDs by MBE[34]. The presence of the 3-fold in-plane periodicity (compared to the 6-fold periodicity for WSe$_2$) is linked to stronger interlayer vdW interactions which is explained in the following sections based on DFT calculations. Hence, Bi$_2$Se$_3$ vdW compounds do not suffer from the fundamental limitation of stacking fault formation in vdW homoepitaxy as generally observed in TMD vdW compounds[21,33,34]. This opens a window for defect-free integration of Bi$_2$Se$_3$ through the growth process of vdW homoepitaxy, and possibly also for other related vdW compounds.

Theoretical DFT calculations are presented for the highlighted 2D chalcogenide materials WSe$_2$ and Bi$_2$Se$_3$ to explain the striking difference in twin defect formation that are observed from the experimental data. In Figure 4, the binding energies are calculated for the set of most stable bilayer stacking configurations for both WSe$_2$ and Bi$_2$Se$_3$. The numerical values of the calculated binding energies are expressed in meV per unit cell and are presented in Figure 4a, with left and right axes corresponding to respectively WSe$_2$ and Bi$_2$Se$_3$. The various stacking configuration are defined by considering a bilayer representation of the atomic layers at the interface, as schematically illustrated in Figure 4b. The stacking configurations are then noted by assigning a letter for each bilayer at the interface in agreement with the void spaces A, B and C, and a prime (') is used when the bonding symmetry of the top bilayer is inversed with respect to the bonding symmetry of the bottom bilayer (see Figure 4b). This notation and bilayer representation enables an appropriate and physically relevant comparison between WSe$_2$ and Bi$_2$Se$_3$ that respectively have a triple- and quintuple-layer structure. The usage of the prime (') easily separates the 0º from the 60º twin, and hence results in six possible stacking configurations for each vdW compound. In Figure 4c, the bond angle (θ), bond length (l) and lattice parameter (a) of both compounds are given to justify the representation used in Figure 4b.

The DFT calculations reveal two important aspects. The first aspect concerns the relative comparison of the most preferred stacking with the most preferred 60º twin for each vdW compound. As seen from Figure 4a, the preferred stacking in WSe$_2$ is noted as AA' and has two stable 60º twins noted as AB and AC. The relative comparison of these configurations reveals that 60º twins in WSe$_2$ (AB and AC) are slightly less stable having a binding energy that is ~2.4 % lower compared to the most preferred stacking (AA'). In the case of Bi$_2$Se$_3$, the most perfect stacking is noted as AB' and similarly has two stable 60º twins (AB and AC). The preference of AB' for Bi$_2$Se$_3$ compared to AA' for WSe$_2$ could be linked to the differences in bond angle, bond length and lattice parameter (Figure 4c). Nevertheless, the relative comparison of these configurations reveals that for Bi$_2$Se$_3$, 60º twins are ~3.4 % less stable compared to the preferred stacking. Hence, both WSe$_2$ and Bi$_2$Se$_3$ only have very subtle differences in binding energy between 0º and 60º twins, which cease to explain the striking difference in twin defect formation as experimentally observed.

The second important aspect that is revealed from our DFT calculations is related to the absolute binding energies of the two vdW compounds. From Figure 4a, it is clearly obvious that the binding energy of the Bi$_2$Se$_3$ compound, in general, is notably larger compared to the binding energy of the WSe$_2$ compound. This can be seen from the absolute values and energy-range difference of the Wse$_2$ and Bi$_2$Se$_3$ axes in Figure 4a. The Bi$_2$Se$_3$ compound generally results in a ~42 % lower (i.e. stronger binding) energy compared to the WSe$_2$ compound. As a result, this notably stronger vdW interlayer coupling in Bi$_2$Se$_3$ explains the striking difference in twin defect formation with respect to WSe$_2$.

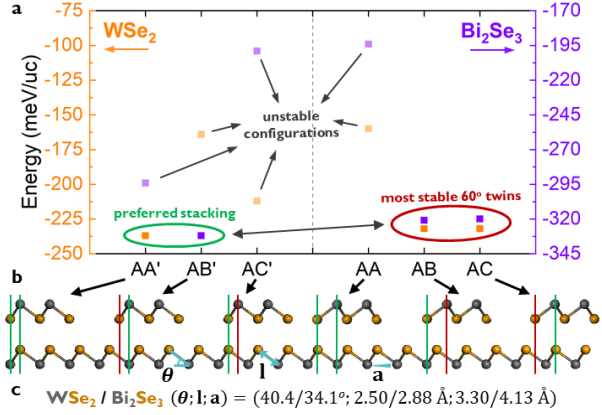

**Figure 4:** Binding energy from DFT calculations for 2D chalcogenides. a) Calculated binding energy in meV per unit cell for both WSe$_2$ and Bi$_2$Se$_3$ bilayer homostructures in function of the bilayer stacking configuration. The WSe$_2$ binding energies are plotted on the left Y-axis (yellow). The Bi$_2$Se$_3$ binding energies - that are significantly larger - are plotted on the right Y-axis (purple). b) Definition, schematic illustration and notation of the various stacking configurations based on a bilayer representation of the atomic layers at the interface. Grey corresponds to the metal atom and yellow to the chalcogen atom. c) Numerical values of the bond angle, bond length and lattice parameter for both WSe$_2$ and Bi$_2$Se$_3$ for the representation used in (b).

The statement made above is confirmed and supported by DFT calculations that consider the alteration of the stacking configuration by nucleus rotation. In Figure 5, the relative energy per unit cell is presented for the stacking alteration from the most preferred stacking configuration to the most preferred 60° twin. In the case of WSe$_2$ (left panel) this is respectively from AA' to AB, and in the case of Bi$_2$Se$_3$ (right panel) this is respectively from AB' to AB. As a result of the stronger interlayer vdW coupling in Bi$_2$Se$_3$, the energy barrier for stacking alteration by nucleus rotation is significantly higher. This is consequently related to the larger amount of energy that is required to rotate the Bi$_2$Se$_3$ nucleus out from its stronger epitaxial registry with the underlying surface. The thermal energy that is available per unit cell to enable this nucleus rotation from 0° to 60° is found insufficient for the case of Bi$_2$Se$_3$ epitaxy ($T_g$ = 160 °C), while it is found sufficient for the case of WSe$_2$ epitaxy ($T_g$ = 450 °C) (see Figure 5). This hence results in a more challenging and hampered rotation of Bi$_2$Se$_3$ vdW nuclei compared to WSe$_2$ vdW nuclei, and confirms the better control on 60° twin defect formation in the case of Bi$_2$Se$_3$ epitaxy.

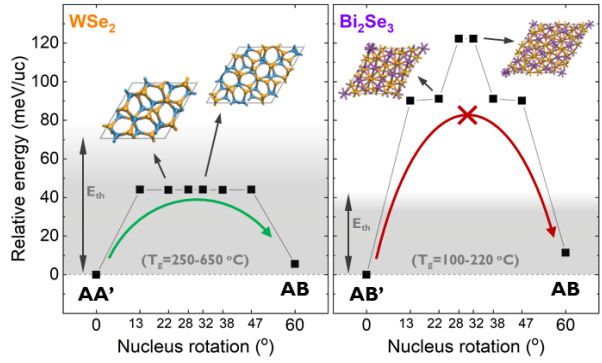

**Figure 5:** Stacking alteration by nucleus rotation from DFT calculations for 2D chalcogenides. Representation of the relative energy in meV per unit cell (left panel WSe$_2$, right panel Bi$_2$Se$_3$) for stacking alteration from the most preferred stacking configuration to the most stable 60° twin. Top-view ball-and-stick schematic illustrations of the intermediated structures at rotation angles of 22° and 32° are illustrated on top of the figure. The thermal energy available per unit cell is shaded in gray for both WSe$_2$ (T$_g$ = 450 °C) and Bi$_2$Se$_3$ (T$_g$ = 160 °C) epitaxy.

In conclusion, it is shown that the formation of 60° twin defects in (quasi-)vdW epitaxy of 2D chalcogenides is significantly reduced through stronger vdW interlayer coupling. In this regard, it is demonstrated that WSe$_2$ quasi vdW heteroepitaxy and vdW homoepitaxy systematically yield a high density of stacking faults resulting from both the weak vdW coupling and the very subtle differences in binding energy for the 0° and 60° configurations. This observation is in striking difference compared to Bi$_2$Se$_3$ quasi-vdW heteroepitaxy and vdW homoepitaxy, where the 3-fold character of the growths reveal the preferred stacking and consequently the significant reduced presence of 60° twin defects. The formation of 60° twins in (quasi-)vdW epitaxy is therefore not only related to the relative difference in binding energy between the 0° and 60° configurations, but also to the absolute strengths of these interlayer vdW interactions. The stronger interlayer coupling in Bi$_2$Se$_3$ compared to WSe$_2$ is shown to challenge the nucleus rotation from the most stable 0° configuration to the most stable 60° twin, which consequently results in twin exclusion in Bi$_2$Se$_3$ epitaxy. The strength of the interlayer vdW coupling in (quasi-)vdW epitaxy is hence a crucial parameter controlling the defect-density of the grown 2D chalcogenides. This opens perspectives for Bi$_2$Se$_3$ and for new vdW materials screening with stronger interlayer coupling to further accelerate the defect-free epitaxial growth of 2D materials.

**ACKNOWLEDGMENTS:**

We acknowledge the Horizon 2020 FETPROAC project SKYTOP – 824123- "Skyrmion - Topological Insulator and Weyl Semimetal Technology".